# Post-Digital Humanities:

## Computation and Cultural Critique in the Arts and Humanities

By David M. Berry

Today we live in computational abundance whereby our everyday lives and the environment that surrounds us are suffused with digital technologies. This is a world of anticipatory technology and contextual computing that uses smart diffused computational processing to create a fine web of computational resources that are embedded into the material world. Thus, the historical distinction between the digital and the non-digital becomes increasingly blurred, to the extent that to talk about the digital presupposes an experiential disjuncture that makes less and less sense. Indeed, just as the ideas of "online" or "being online" have become anachronistic as a result of our always-on smartphones and tablets and widespread wireless networking technologies, so too the term "digital" perhaps assumes a world of the past.

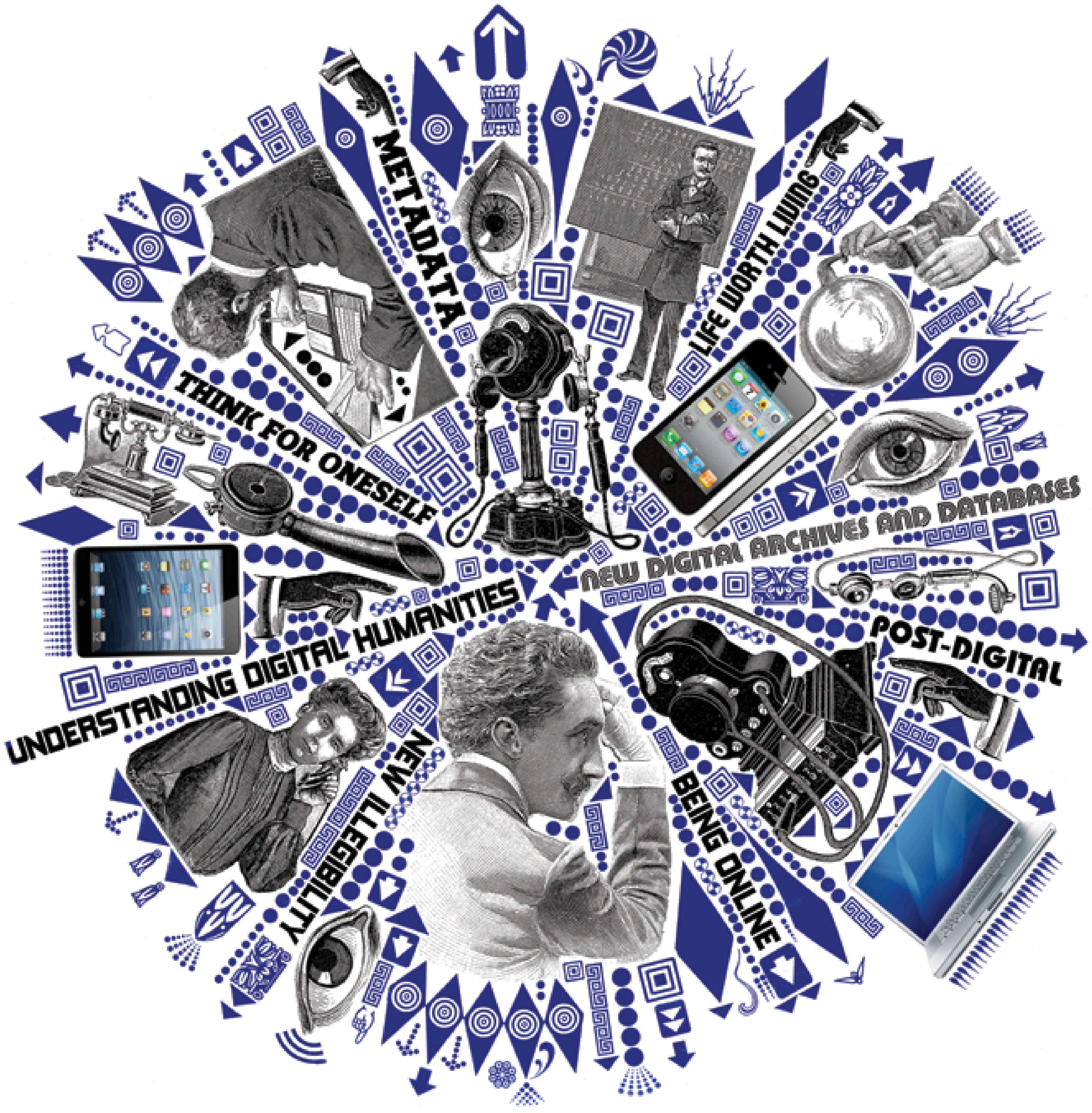

ILLUSTRATION BY LORENZO PETRANTONI, © 2014

As long back as the year 2000, Kim Cascone argued: "The revolutionary period of the digital information age has surely passed. The tendrils of digital technology have in some way touched everyone." He coined the term *post-digital* as a means of thinking about this aesthetic.[1] Indeed, the explosion of digital information and data, combined with the contraction of the time available to deal with the information and data, has created a tendency to understand the "digital" as a "spectatorial condition" whereby "we filter and graze, skim and forward." Similar to the art world, where "mainstream contemporary art simultaneously disavows and depends on the digital revolution,"[2] mainstream humanities research today equally disavows and depends on the digital—to the extent that to ask the question of the distinction raised *by* the digital *for* the humanities is a question that the digital humanities has sought to address.[3]

The digital humanities, at its most straightforward, is the application of computational principles, processes, and machinery to humanities texts—and here I use "texts" to refer to all forms of materialized cultural forms such as images, books, articles, sound, film, video, and so on. Indeed, as I noted in my introduction to *Understanding Digital Humanities,* the digital humanities tries "to take account of the plasticity of digital forms and the way in which they point towards a new way of working with representation and mediation, what might be called the digital 'folding' of memory and archives, whereby one is able to approach culture in a radically new way."[4] Much of the early work of the digital humanities was focused on getting traditional humanities materials into a form whereby they could be subject to computational work—through digitalization projects, new digital archives and databases, and the "marking up" of texts to enable computational analysis. However, the digital humanities has also had to come to terms with new forms of digital collections and archives, such as the web itself and the archives made from it (e.g., the Internet Archive), and indeed with archives and databases that may be made up of data about data, so called *metadata,* and computational non-human-readable materials.

Thus we enter a time of a new illegibility: we might say that we can no longer read what we are writing; we increasingly rely on digital technology both to write and to read for us as a form of algorithmic inscription. We not only are facing these new forms of grammatization but also are entering a phase whereby the grammatization process produces symbols and discrete representational units that become opaque to us even as they are drawn from us through technical devices that monitor and track us. As Bernard Stiegler explains, digital technology engenders "a process of the grammatization of flows, as a process of discretization—for example, of the flow of speech, or the flow of gestures of the worker's body—[this] is what makes possible... technical reproducibility and thus... control, and therefore, by the same stroke, the organization of short circuits inside the long circuits constitutive of transindividuation."[5]

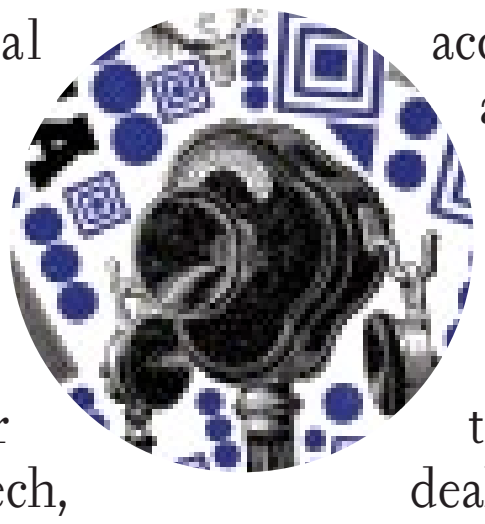

This process of transindividuation, through practices such as a humanities education, creates the psychic structures for the possibility of thinking. These structures are constitutive of the development of the "maturity" of the individual and the collective formation and education of intelligence and knowledge. It is through transindividuation, Stiegler argues, that the ability to "think for oneself" is developed and one can begin to outline what is a "life worth living," a concern to which the humanities have traditionally been linked. According to Stiegler, it is in this destabilizing and deconstructing moment that the digital, as presently constructed, also *undermines* the development of attention, memory, concentration, and intelligence.[6]

Indeed, the question the digital poses to the humanities is addressed directly at what Imre Lakatos would have called the "hard core" of the humanities, the unspoken assumptions and ontological foundations that support the "normal" research undertaken by humanities scholars on an everyday basis—for example, the notion of a hermeneutically accessible "text" as a constitutive and foundational concept.[7] The digital humanities has attempted to address these issues with notions of "close" and "distant" reading, particular practices related to dealing with both small and larger numbers of texts. However, the digital humanities remains somewhat ill-equipped to deal with the hermeneutic challenges of computer-generated data, which nonetheless contains some sense of internal structure, meaning, and in some instances, narrative but which is structured in "chains" that are not conducive to human memory and understanding. Indeed, this raises the question of what the research programs relevant to a post-digital humanities might be—a question of both research and practice, theoretical work and building things, technologically engaged work and critical technical practice.

At the same time, and from a different direction, digital technologies have undermined and reconfigured the very texts that humanities and digital humanities scholars have taken as their research objects; digital technologies have re-presented these texts as fragmentary forms, often realigned and interleaved with fragments from other texts. This interdiscursivity and intertextuality of the digital has, of course, been much remarked upon and even used creatively in the writing of new forms of digital and e-literature. However, in a new way, this process has, to follow Stiegler, begun to undermine the "long circuits" of culture, such that we no longer describe a method, such as the production of concordances within the digital humanities, but instead describe a logic of computational media from which no "long chains" are reconstructed from their shorter counterparts. Stiegler views this as a serious danger to societies as they deconstruct the very structures of education and learning on which they are built. Indeed, he calls for the creation of counter-products that might reintroduce singularity into cultural experience and somehow disconnect desire from the imperatives of consumption.

**Might it be better to raise the question of the "post-digital," since we are rapidly entering a moment when the difficulty will be found in encountering culture outside of digital media?**

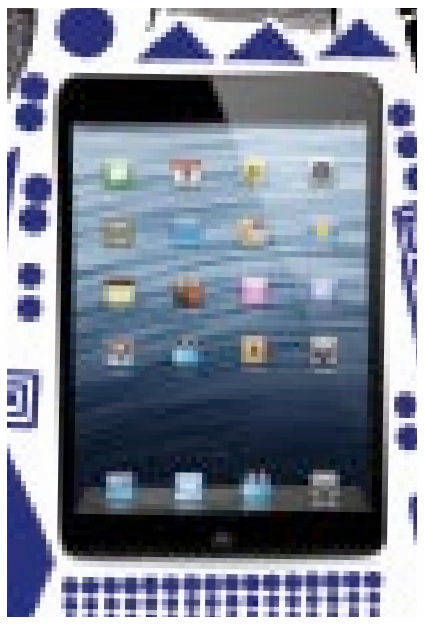

In which case, should we ask about a post-digital humanities that is adequate to begin to address this problem? In other words, in a time of computational abundance, might it therefore be better to raise the question of the "post-digital," since we are rapidly entering a moment when the difficulty will be found in encountering culture outside of digital media? As Florian Cramer argues: "In a post-digital age, the question of whether or not something is digital is no longer really important—just as the ubiquity of print, soon after Gutenberg, rendered obsolete all debates (besides historical ones) about the 'print revolution.'"[8] This is to move away from a comparative notion of the digital—contrasted with other material forms such as paper, celluloid, or photopaper—and instead begin to think about how the digital is modulated within various materialities.

Thus, the post-digital is represented by and indicative of a moment when the computational has become hegemonic. The digital is then understood as a historic moment defined in opposition to the analog. We might no longer talk about digital versus analog but instead about modulations of the digital or different intensities of the computational. We should therefore critically analyze the way in which cadences of the computational are made and materialized.

The post-digital humanities would then be attentive to our massively computational world and culture, as well as to the ways in which culture is materialized and fixed in forms specific to digital material culture—that is, to how culture is inscribed not just in moments of culture created by human actors but also in the technical devices, recording systems, trackers, web bugs, and beacons of the digital age, together with the giant databases they fill with narratives and documentaries about the great and the mundane, the event and the everyday. The post-digital humanities would then be attentive to the way in which human culture, writ large, is also written digitally, in an open-ended arrangement of diverse practices and parts. This would be a digital humanities that includes cultural critique, as called for by Alan Liu.[9] Indeed, critique is a necessary precondition for asking the kinds of questions that the post-digital raises in relation to issues of power, domination, myth, and exploitation, but also in relation to the historical, social, political, and cultural contexts that cadences of the digital make possible today. ■

**Notes**

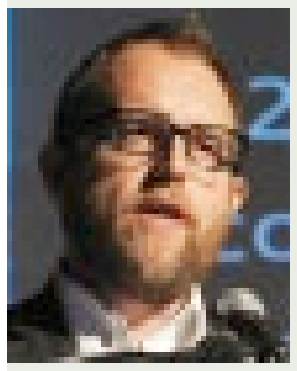

**David M. Berry** (D.M.Berry@sussex.ac.uk) is Reader (Associate Professor) in Media and Communication and is Co-Director of the Centre for Material Digital Culture at the University of Sussex. He blogs at *Stunlaw: A Critical View of Politics, Arts, and Technology* (http://stunlaw.blogspot.com/).